\documentclass{article}
\usepackage[preprint]{spconf}
\usepackage{amsmath,graphicx}
\usepackage{algorithm}
\usepackage{algorithmic}
\usepackage{booktabs}
\usepackage{microtype}
\usepackage{graphicx}
\usepackage{subfigure}
\usepackage{hyperref}
\usepackage{multirow}
\usepackage{color}
\usepackage{amsthm}
\usepackage{amssymb}
\usepackage{mathrsfs}
\usepackage{xcolor}
\usepackage{dsfont}

\copyrightnotice{978-1-6654-7189-3/22/\$31.00~\copyright2023 IEEE}

\title{SVLDL: Improved Speaker Age Estimation Using Selective Variance Label Distribution Learning}
\name{Zuheng Kang, Jianzong Wang*\thanks{*Corresponding author: Jianzong Wang, jzwang@188.com}, Junqing Peng, Jing Xiao}
\address{Ping An Technology (Shenzhen) Co., Ltd.}
\begin{document}
%
\maketitle

\begin{abstract}
    Estimating age from a single speech is a classic and challenging topic.
    Although Label Distribution Learning (LDL) can represent adjacent indistinguishable ages well, the uncertainty of the age estimate for each utterance varies from person to person, i.e., the variance of the age distribution is different.
    To address this issue, we propose selective variance label distribution learning (SVLDL) method to adapt the variance of different age distributions.
    Furthermore, the model uses WavLM as the speech feature extractor and adds the auxiliary task of gender recognition to further improve the performance.
    Two tricks are applied on the loss function to enhance the robustness of the age estimation and improve the quality of the fitted age distribution.
    Extensive experiments show that the model achieves state-of-the-art performance on all aspects of the NIST SRE08-10 and a real-world datasets.
\end{abstract}

\begin{keywords}
    speaker age estimation, label distribution learning, multi-task learning, gender recognition
\end{keywords}

\section{Introduction}
Speech is the sound produced by the accurate coordinated movement of multiple organs in the human body.
Hence, the acoustic characteristics of speech can transmit information about the physical characteristics of the speaker.
The rapid development of new speech applications requires techniques capable of estimating information on various biological attributes of such speakers.
Recently, deep-learning-based approaches show great performance in extracting hidden speech information, including facial expression \cite{si2021speech2video} and emotion \cite{kang2022speecheq}, and age \cite{si2022towards}, etc.
If such speech features can be used to automatically estimate a speaker's age, it could be widely used for human-computer interaction, forensics, and other purposes.

Many researchers have studied the performance of human and artificial intelligence systems in estimating age from speech.
The results show that the average error of humans judging the age of adults is about 10 years old, and the judgment of the age of children is about 1-year old \cite{huckvale2015comparison}.
The performance of age estimates may also have implications for human development.
\cite{shivakumar2021phone} collected the speech of children.
It can be seen that, as children gradually enter puberty, changes in the vocal cords can affect age estimates and increase uncertainty.
In adulthood, the vocal cords are fully developed and the change tends to be slow.
However, as we age, various organs experience regular aging: the voice changes from bright to hoarse, and articulation from clear to vague \cite{tucker2021speech,kitagishi2020speaker}.
Judgments at different ages also have different uncertainties, and these uncertainties may vary from age to age, from utterance to utterance.

\begin{figure*}[t]
    {\includegraphics[width=0.98\textwidth]{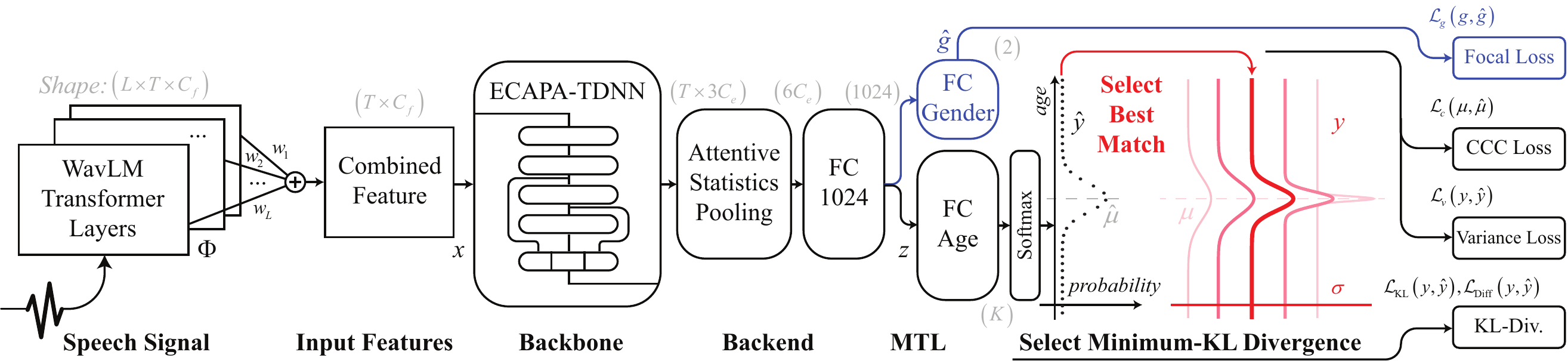}}
    \centering
    \caption{Network topology of the SVLDL framework.
        ``FC'' denotes a fully connected layer.
        $ \oplus $ denotes element-wise addition.}
    \label{fig:overview}
\end{figure*}

Traditional methods for speaker age estimation can be generally classified into classification-based and regression-based methods.
Most researchers mainly focus on the exploration of backbone model structures, such as deep neural network (DNN) \cite{kalluri2019deep}, i-vector \cite{bahari2014speaker}, x-vector \cite{ghahremani2018end,kwasny2021explaining} or adding attention mechanism \cite{kaushik2021end}.
Some researchers have tried different machine learning features, such as the OpenSmile toolbox \cite{eyben2010opensmile} to study this problem \cite{shaqra2019recognizing,burkhardt2021age}.
As manipulated acoustic features, such as mel-filter banks, encounter performance bottlenecks, some researchers use other speech features for modeling, which can capture acoustic features that are imperceptible to the human ear, such as SincNet \cite{pan2021multi} take full advantage of acoustic information, resulting in improved performance.
However, these features are only direct translations of speech signals, not language models for understanding human speech.
Self-supervised learning (SSL) generates high-quality speech features with language model (such as wav2vec \cite{baevski2020wav2vec} and WavLM \cite{chen2022wavlm}) by learning from a large amount of data \cite{hendrycks2019using}.
By injecting this prior knowledge, speech age estimation achieves better performance \cite{gupta2022estimation}.
Although these methods have achieved great results, they ignored the fact that it rarely considers the relationship between labels, such as order and adjacent correlations, which are important clues for speaker age estimation.
Since speaker age labels form an ordered set of numbers, significant ordinal relationships and adjacencies between labels should be fully exploited to achieve higher performance.

Label distribution learning (LDL) \cite{geng2016label} addresses the above problems by transforming the classification problem into a distribution learning task that minimizes the difference between the predicted and constructed Gaussian distributions of labels.
In the field of computer vision, impressive progress has been made in facial age estimation, where LDL shows great potential \cite{zhang2021practical}.
Framework \cite{si2022towards} applied this method to the speaker age recognition task and achieved good performance.
Since the uncertainty of each person is different, i.e., the variance of the Gaussian distribution varies from person to person, adaptive-based LDL methods have been proposed successively \cite{geng2014facial, wen2020adaptive, li2022unimodal}.
However, loss functions that measure regression error often use simple metrics, such as L1 distance, which are not dynamically adjusted for a specific distribution at training time.
This method does not achieve optimal regression performance.
Meanwhile, these algorithms do not get the correct shape of the learned distribution, which may lead to multimodal problems (multiple peaks in the fitted distribution).

Additionally, Multi-task learning (MTL) uses a shared backbone model to simultaneously optimize objectives for different tasks.
The advantage comes from adding more useful information while optimizing the original model.
In speaker age estimation, adding the task of gender recognition has been shown to improve performance \cite{gupta2022estimation, huynh2020joint}.
Meanwhile, in regression problems, Lin's consistent correlation coefficient loss \cite{lawrence1989concordance} also achieves a lot of performance gains by replacing L1 or L2 distance-based losses.

Considering the above advantages and disadvantages, we have made the following improvements and contributions:

\noindent
\begin{itemize}
    \item We improve the original label distribution learning (LDL) method and propose a new selective variance label distribution learning (SVLDL) method that adaptively selects the optimal distribution that matches the variance.
    \item The quality of fitted distributions is improved by fitting additional first-order difference distribution, and a brief theoretical proof is given.
    \item The age estimation performance is enhanced by using Lin's concordance correlation coefficient \cite{lawrence1989concordance} loss.
    \item The performance was improved by adding an auxiliary task for gender recognition and using WavLM as the speech feature extractor.
    \item Experimental results on the publicly available NIST SRE08-10 dataset and a real-world dataset show that the improved SVLDL framework achieves state-of-the-art performance compared to the framework \cite{si2022towards}.
\end{itemize}

\section{Methodology}

\subsection{Network Architecture}

Figure \ref{fig:overview} outlines the pipeline of the proposed method.
Since the structure of ECAPA-TDNN \cite{desplanques2020ecapa} has an efficient design structure, such as Res2Net \cite{gao2019res2net} and squeeze excitation blocks (SE) \cite{hu2018squeeze}, it is used as the backbone model.
All the information on the time dimension is collected through attentive statistics pooling (SP).
After the SP, there are two fully connected layers, and finally a softmax layer is connected to obtain the output distribution of the labels, denoted as $ y $; the output of the middle layer is denoted as $ z $, which is also used as input for the auxiliary task of gender recognition.

\subsection{Self-supervised Representation}

Motivated by the successful application of self-supervised learning (SSL) in various speech domains, we explore the use of WavLM \cite{chen2022wavlm} on the task of speaker age estimation.
The WavLM model learns speech representations by solving contrastive tasks in a latent space in a self-supervised manner.
It tries to recover the randomly masked part of the encoded audio features.
By learning from large amounts of real multilingual, multi-channel unlabeled data, SSL models can deeply understand contextual information and produce high-quality speech representations in the latent space.

In our framework, seen from Figure \ref{fig:overview}, we utilize all latent output of WavLM transformer layers $ \Phi =\left( \phi _1,...,\phi _L \right) $ and assign a trainable weight $ W=\left( w_1,...,w_L \right) $ to each of them.
The weighted sum is then used to generate speech features $ x=\sum_{i=1}^L{\left( \phi _i\cdot w_i \right)} $, where $ \Phi \in \mathbb{R}^{L\times T\times C_f} $, $ x\in \mathbb{R}^{T\times C_f} $, $ T $ is number of time frames, $ C_f $ is the feature size, $ L $ is the number of layers of WavLM.
In this way, the model can make full use of speech information from shallow to deep, from concrete to abstract.

\subsection{Label Distribution Learning}

Before introducing SVLDL, we need to know how LDL works and understand some parameters, $ \hat{\mu}_n $ and $ \hat{\sigma}_n $ are the mean and standard deviation of the predicted distribution, and $ \mu_n $ and $ \sigma_n $ are for the ground-truth of sample $ n $.
Where $ \hat{\mu}_n=\sum_{k=1}^K{k\cdot \hat{y}_{n}^{k}} $, and $ \hat{\sigma}_n=\frac{1}{N}\sum_{k=1}^K{\left( \hat{y}_{n}^{k}-\hat{\mu} _n \right) ^2} $.
To take advantage of the intrinsic relationship in model outputs, we treat these outputs as a distribution representing the predicted age distribution.
The label distribution $ \hat{y}_{n}^{k} $ is a predicted probability distribution, which satisfy $ \hat{y}_{n}^{k}\in \left[ 0,1 \right] $ and $ \sum_{k=1}^K{\hat{y}_{n}^{k}}=1 $, $ n $ is the data sample, $ k $ is the age label, $ k \in [1, K] $ and $ K $ denotes the maximum age.
In age estimation, age is usually represented using a Gaussian distribution centered around the ground-truth age $ \mu _n $.
This ground truth probability distribution $ y_{n}^{k} $ is represented by a Gaussian distribution function.

\noindent
\begin{equation}
    \small
    y_{n}^{k}=C_n\cdot e^{-\left( k-\mu _n \right) ^2/\left( 2\sigma ^2 \right)}
    \label{eq:gaussian}
\end{equation}
\noindent

where $ \sigma $ is a fixed value that is reasonably chosen in LDL, $ C_n $ is a constant to make $ \sum_k{y_{n}^{k}=1} $.
The difference between the ground truth label distribution $ \hat{y} $ and the predicted distribution $ y $ is measured using the Kullback-Leibler divergence (KL divergence).
Therefore, the loss function $ \mathcal{L}_{\text{KL}} $ can be defined as,

\noindent
\begin{equation}
    \small
    \mathcal{L}_{\text{KL}}\left( y,\hat{y} \right) =\frac{1}{N}\sum_{n=1}^N{\text{D}_{\text{KL}}\left( y_n|\hat{y}_n \right)}=\frac{1}{N}\sum_{n=1}^N{\sum_{k=1}^K{y_{n}^{k}\log \left( \frac{y_{n}^{k}}{\hat{y}_{n}^{k}} \right)}}
    \label{eq:loss_kl}
\end{equation}
\noindent

where $ N $ denotes number of data samples, and $ y_n=\left( y_{n}^{1},...,y_{n}^{k} \right) $ and $ \hat{y}_n=\left( \hat{y}_{n}^{1},...,\hat{y}_{n}^{k} \right) $.
However, not all predicted distributions need to follow the same variance $ \sigma $, that is, the value of $ \sigma $ needs to be chosen adaptively for each utterance.

\subsection{Selective Variance Label Distribution Learning}
According to the principles discussed earlier, the age distribution of learning should vary by the utterance.
To achieve this goal, we propose a novel selective variance label distribution learning method that fully adapts to the variance of each utterance.
That is, the process of selecting the best matching distribution from a series of red candidate distributions, shown in Figure \ref{fig:overview}.
These candidate Gaussian distributions can be defined as,

\noindent
\begin{equation}
    \small
    y_{n}^{k}\left( s \right) =C_n\cdot e^{-\left( k-\mu _n \right) ^2/s}
    \label{eq:gaussian_svldl}
\end{equation}
\noindent

Where $ s \in S $, and $ S $ is a set of predefined candidate variance values.
Among the candidate distributions obtained using these values, there should be one that matches the ground-truth age distribution as closely as possible.
Therefore, the problem turns into choosing the smallest difference between a set of candidate label distributions $ \left[ \hat{y} \right] $ and the predicted distribution $ y $.
Denote that $ s^* $ is the variance of the best matching case, which is related to the ground-truth value of standard deviation with $ s^*=\sigma _{n}^{2} $.
Then the loss function $ \mathcal{L}_{\text{KL}} $ as follows,

\noindent
\begin{equation}
    \small
    \begin{aligned}
        \mathcal{L}_{\text{KL}}\left( y,\hat{y} \right) & =\frac{1}{N}\sum_{n=1}^N{\left( \underset{s\in S}{\text{arg}\min}\left( \text{D}_{\text{KL}}\left( y_n\left( s \right) |\hat{y}_n \right) \right) \right)}
        \\
                                                        & =\frac{1}{N}\sum_{n=1}^N{\text{D}_{\text{KL}}\left( y_n\left( s^* \right) |\hat{y}_n \right)}
    \end{aligned}
    \label{eq:loss_kl_svldl}
\end{equation}
\noindent

In this way, the algorithm can adaptively select the best matching variance of the Gaussian distribution for training.

\subsection{Unimodal Distribution Constraints}

In experiments, we observe that the baseline based on mean-variance learned distributions is multimodal for some instances, in Figure \ref{fig:multimodal}.
Namely, there will be multiple peaks in the distribution.
We propose an approach to overcome this issue by simultaneously learning the first-order differences of the distributions.
Suppose $ \Delta \left( \cdot \right) $ is the first-order difference function of a discrete distribution, in Equation \ref{eq:loss_kl_svldl} with variance $ s^* $.
This method is denoted as Diff.
The loss function $ \mathcal{L}_{\text{Diff}} $ used to optimize the first-order difference of the distribution is as follows,

\noindent
\begin{equation}
    \small
    \begin{aligned}
        \mathcal{L}_{\text{Diff}}\left( y,\hat{y} \right) & =\frac{1}{N}\sum_{n=1}^N{\left( \Delta \left( y_n\left( s^* \right) \right) -\Delta \left( \hat{y}_n \right) \right) ^2}
        \\
                                                          & =\frac{1}{N}\sum_{n=1}^N{\sum_{k=1}^{K-1}{\left( \Delta \left( y_{n}^{k}\left( s^* \right) \right) -\left( \hat{y}_{n}^{k+1}-\hat{y}_{n}^{k} \right) \right) ^2}}
    \end{aligned}
    \label{eq:loss_diff}
\end{equation}
\noindent

\textbf{Proof:}
To demonstrate that Equation \ref{eq:loss_diff} constrains the distribution to be unimodal, assuming that the first difference of distribution $ \Delta \left( y_{n}^{k} \right) $ is proportional to the first derivative of the distribution $ y'_n=\text{d}y_{n}^{k}/\text{d}k $.
Since $ y $ is a Gaussian distribution, when $ k<\mu_n $, $ y'_n>0 $, and when $ k>\mu_n $, $ y'_n<0 $.
In order to show how our loss to be unimodal, we take a case of $ k<\mu_n $ for illustration, where $ \hat{y}_{n}^{k+1}-\hat{y}_{n}^{k}>0 $.
If it is not a unimodal case at $ \hat{y}_{n}^{k+1}-\hat{y}_{n}^{k}<0 $, to verify that the loss function $ \mathcal{L}_\text{Diff} $ can constrain the distribution to a single mode, we calculate the gradient of this loss function over $ \hat{y}_{n}^{k} $ and $ \hat{y}_{n}^{k+1} $ respectively.

\noindent
\begin{equation}
    \small
    \frac{\partial \mathcal{L}_{\text{Diff}}}{\partial \hat{y}_{n}^{k}}\propto 2\left( y'_n-\left( \hat{y}_{n}^{k+1}-\hat{y}_{n}^{k} \right) \right) >0
    \label{eq:partial_k}
\end{equation}

\noindent
\begin{equation}
    \small
    \frac{\partial \mathcal{L}_{\text{Diff}}}{\partial \hat{y}_{n}^{k+1}}\propto -2\left( y'_n-\left( \hat{y}_{n}^{k+1}-\hat{y}_{n}^{k} \right) \right) <0
    \label{eq:partial_k_1}
\end{equation}
\noindent

According to Equations \ref{eq:partial_k} and \ref{eq:partial_k_1}, $ \hat{y}_{n}^{k} $ decreases due to its positive gradient and $ \hat{y}_{n}^{k+1} $ increases due to its negative gradient.
In addition, the magnitude of this gradient is taken from the first-order difference of the Gaussian distribution, so the loss function can better constrain the distribution to the shape of the Gaussian distribution, thereby improving the quality of the fitted distribution.

\subsection{Hybrid Loss}

For regression predicting age, the Lin's Concordance Correlation Coefficient (CCC) \cite{lawrence1989concordance} is more reliable to use, denoted as $ \rho _c $.
CCC is a measure of the agreement between ground-true labels and predicted labels.
If the predicted value changes, the score is proportional to its deviation \cite{atmaja2021evaluation}.
\cite{pandit2019many} provides complete proof that CCC outperforms other common regression losses, and we will use experiments to verify that it also holds for speech age estimation.
Therefore, the loss function $ \mathcal{L}_\text{c} $ derived from $ \rho _c $ is used as a measure of regression age,

\noindent
\begin{equation}
    \small
    \mathcal{L}_{\text{c}}\left( \mu ,\hat{\mu} \right) =1-\rho _c=1-\frac{2\sigma _{\left[ pt \right]}^{2}}{\sigma _{\left[ p \right]}^{2}+\sigma _{\left[ t \right]}^{2}+\left( \mu _{\left[ p \right]}-\mu _{\left[ t \right]} \right) ^2}
    \label{eq:ccc}
\end{equation}
\noindent

Where $ \sigma _{\left[ pt \right]}^2=\text{cov}\left( \mu ,\hat{\mu} \right) $, $ \mu _{\left[ p \right]}=\mathbb{E}\left( \hat{\mu} \right) $, $ \mu _{\left[ t \right]}=\mathbb{E}\left( \mu \right) $, $ \sigma _{\left[ p \right]}^{2}=\text{var}\left( \hat{\mu} \right) $, $ \sigma _{\left[ t \right]}^{2}=\text{var}\left( \mu \right) $, w.r.t. $ n $.

Since human voice aging is a continuous process, the predicted age should be more likely to be the ground-truth age, and the farther away from this age, the less likely it is.
Smaller variance means lower uncertainty in age prediction.
The variance loss $ \mathcal{L}_\text{v} $ reduces the uncertainty in the estimated age distribution,

\noindent
\begin{equation}
    \small
    \mathcal{L}_{\text{v}}\left( y,\hat{y} \right) =\frac{1}{N}\sum_{n=1}^N{\sum_{k=1}^K{\left( \hat{y}_{n}^{k}\cdot \left( k-\hat{\mu}_n \right) ^2 \right)}}
    \label{eq:loss_v}
\end{equation}
\noindent

Due to the physiological differences between men and women, there are distinct differences in speech characteristics -- the average formant and fundamental frequency of women's speech sounds higher than those of men \cite{wu1991gender, mendoza1996differences}.
By using a multi-task learning approach while performing gender recognition tasks, gender information will be implicitly added to the model (blue task in Figure \ref{fig:overview}).
In this task, the gender classification task is trained with a focal loss (FL) \cite{lin2017focal} with a tunable focus parameter $ \gamma \ge 0 $.
The loss function $ \mathcal{L}_\text{g} $ is defined as follows,

\noindent
\begin{equation}
    \small
    \mathcal{L}_g\left( g,\hat{g} \right) =\text{FL}\left( g,\hat{g} \right)
    \label{eq:fl}
\end{equation}
\noindent

Where $ \hat{g} $ and $ g $ are the predicted and the ground-truth gender.
The overall loss is that given in Equation \ref{eq:loss}, where $ \lambda _1 $, $ \lambda _2 $, $ \lambda _3 $, $ \lambda _4 $ and $ \lambda _5 $ are hyper-parameters.

\noindent
\begin{equation}
    \mathcal{L}=\lambda _1\cdot \mathcal{L}_{\text{c}}+\lambda _2\cdot \mathcal{L}_{\text{KL}}+\lambda _3\cdot \mathcal{L}_{\text{v}}+\lambda _4\cdot \mathcal{L}_{\text{Diff}}+\lambda _5\cdot \mathcal{L}_{\text{g}}
    \label{eq:loss}
\end{equation}
\noindent

\subsection{Training and Inference}

During the training phase, speech activity detection (SAD) preprocesses the audio to remove non-speech frames since speech may contain many silent segments.
In our experiments, the rvad model \cite{tan2020rvad} is used for this task.
In order to make full use of hardware resources to train models quickly, model training can be divided into two stages.

\noindent
\textbf{Offline training:}
Since the inference speed of the WavLM model is not fast, first convert all the data in the dataset into speech features and save them in Numpy format, and then use these converted data directly to speed up training.

\noindent
\textbf{Online training:}
To improve the robustness of the model, we employ a chain-like augment:
(1) Noise was added using MUSAN.
(2) The RIR reverb is added.
(3) Time stretch augment \cite{park2019specaugment}: time stretching doesn't change pitch, it simulates a person's different speech rates.

During the inference phase, the age estimate of the utterance and its uncertainty are the mean age $ \hat{\mu} $ and variance $ \hat{\sigma} $ of the predicted distribution.
At the same time, the auxiliary task of gender recognition will be abandoned.
The speech will be processed by SAD first, and then the whole segment will be sent to the model for prediction.

\section{Experiments}

\subsection{Datasets}

To demonstrate the advantages of the proposed method, we use the same dataset and conduct experimental validation under the same settings as \cite{si2022towards}.

\noindent
\textbf{NIST SRE08-10 dataset:}
We use 11,205 utterances (458 male and 769 female speakers) from NIST SRE08 as the training set, and 5,331 telephone-conditioned utterances (236 male and 256 female speakers) from NIST SRE10 as the test set, similar to \cite{sadjadi2016speaker}.
The speech in the dataset contains both English and non-English.
Neither the speakers nor the recordings in the training and test set overlap.
The speech in the dataset contains Chinese and Chinese dialects.

\noindent
\textbf{Real-world PA-Age Dataset:}
This dataset is from the financial insurance domain and contains 69,610 corpora (30,661 male and 28,386 female corpora).
The test set used the same 4,000 utterances as \cite{si2022towards}.
The average duration of effective speech is 28.125 seconds, and the standard deviation of duration is 19.114 seconds.

\subsection{Metrics}

To evaluate how good our age estimator is, we report regression performance in terms of mean absolute error (MAE) and Pearson's correlation coefficient (PCC) $ \rho $.
It is defined in Equation \ref{eq:mae} and \ref{eq:pcc}.
The lower MAE and higher PCC, the better.

\noindent
\begin{equation}
    \small
    \text{MAE}=\frac{1}{N}\sum_{n=1}^N{\lVert \hat{\mu}_n-\mu _n \rVert _1}
    \label{eq:mae}
\end{equation}
\noindent

\noindent
\begin{equation}
    \small
    \rho =\frac{1}{N-1}\sum_{n=1}^N{\left( \left( \frac{\hat{\mu}_n-\mu _{\left[ p \right]}}{\sigma _{\left[ p \right]}} \right) \left( \frac{\mu _n-\mu _{\left[ t \right]}}{\sigma _{\left[ t \right]}} \right) \right)}
    \label{eq:pcc}
\end{equation}
\noindent

To measure whether the resulting distribution is unimodal, we introduce a unimodal coefficient $ \eta _q $, representing the average number of modes within $ q $ standard deviations which can roughly detect the number of peaks in the predicted distribution, in Equation \ref{eq:unimodal}.
The lower the better.
In order to ensure the accuracy of the detection, it is only necessary to consider the age within $ q $ standard deviations ($ q=2 $ in experiment).

\noindent
\begin{equation}
    \small
    \eta _q=\frac{1}{N}\sum_{n=1}^N{\left( \sum_{k_{\min}<k<k_{\max}}{\mathds{1}\left( \text{cond}\left( n,k \right) \right)} \right)}
    \label{eq:unimodal}
\end{equation}

\noindent
\begin{equation}
    \small
    \text{cond}\left( n,k \right) =\left( \Delta \left( \hat{y}_{n}^{k} \right) <0 \right) \land \left( \Delta \left( \hat{y}_{n}^{k+1} \right) >0 \right)
    \label{eq:unimodal_cond}
\end{equation}

\noindent
\begin{equation}
    \small
    \begin{aligned}
        k_{\min} & =\max \left( 1,\hat{\mu}_n-q\cdot \hat{\sigma}_n \right)   \\
        k_{\max} & =\min \left( \hat{\mu}_n+q\cdot \hat{\sigma}_n,K-1 \right)
    \end{aligned}
    \label{eq:unimodal_k_range}
\end{equation}
\noindent

Where $ \mathds{1}\left( \cdot \right) $ is a function that converts a boolean value to an integer.
Equation \ref{eq:unimodal_cond} is the conditional function to detect peaks in the distribution.
Equation \ref{eq:unimodal_k_range} defines the age range to be calculated.

\subsection{Hyper-parameters}

For speech features, the WavLM model uses the ``WavLM Base+'' setting in our implementation, which has 13 transformer encoder layers \cite{vaswani2017attention}, 768-dimensional hidden states, and 8 attention heads.
The channel parameter $ C_b $ for ECAPA-TDNN in the convolutional layers for the proposed network is 256.
The bottleneck size in the SE-Block and Attention modules is set to 128.
The scale size in Res2Block is set to 8.
The tunable parameter $ \gamma $ in FL is 10.
The maximum age label $ K $ is set to 100.

The model is first quickly trained by offline training and then fine-tuned by data augmentation using online training.
During the offline training phase, the SGD \cite{bottou2012stochastic} optimizer uses a momentum of 0.9 and weight decay of 1e-3.
The mini-batch is set to 64 and the initial learning rate is 2e-3 to train all our models.
The speech features are segmented into about 3 seconds (150 WavLM frames) to avoid over-fitting and to speed up training.
The set of standard deviation candidates $ S $ are: from 0.01 to 10 in steps of 0.1, i.e., $ S=\left\{ 0.1,0.2,...,10 \right\}^2 $.
In the fine-tuning stage with online training mode, the speech segment changed to 6 seconds.
Due to hardware resource constraints, we adopted a smaller fine-tuning learning rate of 1e-5, the weight decay becomes 4e-4, and a batch size of 128.
The values of standard deviation candidates $ S $ have become finer: from 0.01 to 10 in steps of 0.01, i.e., $ S=\left\{ 0.01,0.02,...,10 \right\}^2 $.

\subsection{Implementation Details}

To verify the effectiveness of our proposed method, we compare it with our last results as a baseline method (results directly copied from \cite{si2022towards}).
The best-performing model uses the ResNet-18 model as the backbone.
The age label distribution is optimized with the mean and variance and the KL divergence of the distribution (denoted as MVKL in Table \ref{tab:result}).

To demonstrate that the model based on WavLM combined with ECAPA-TDNN outperforms the previous backbone model, we only replace this part and conduct experiments with the same parameters and methods as before based on optimizing LDL and MVKL (with a fixed variance $ \sigma=1 $).
To show that the SVLDL and CCC are of great help to the age estimation, let $ \lambda _1=10$, $\lambda _2=1 $, $ \lambda _3=0.1 $ and $ \lambda _4=\lambda _5=0 $ for experiments (denoted as CVKL in Table \ref{tab:result}).
To eliminate the multimodality of the age distribution, we give a weight to the loss function of the KL distance of the first-order difference distribution to eliminate this effect, let $ \lambda _1 $, $\lambda _2 $, $ \lambda _3$, $ \lambda _5 $ are the same, $\lambda _4=0.1 $.
In addition, by adding the auxiliary task of gender recognition, the model can improve the discrimination of gender, thereby improving the performance of the age recognition task, let $ \lambda _5=0.01 $, rest are the same.

\subsection{Evaluation Results}

\begin{table}[ht]
    \centering
    \small
    \setlength{\tabcolsep}{6pt}
    \renewcommand{\arraystretch}{1.0}
    \caption{Ablation study on our proposed system compared with baseline model.
        + denotes stacking our methods.}
    \begin{tabular}{@{}lllllll@{}}
        \toprule
                    & \multicolumn{3}{c}{SRE08-10} & \multicolumn{3}{c}{PA-Age}                                                                 \\ \cmidrule(l){2-7}
                    & MAE                          & $ \rho $                   & $ \eta _2 $   & MAE           & $ \rho $      & $ \eta _2 $   \\ \midrule
        \multicolumn{7}{l}{\textbf{\textit{ResNet-18}} \cite{si2022towards}}                                                                    \\
        +LDL+MVKL   & 4.62                         & 0.87                       &               & 6.23          & 0.82          &               \\ \midrule
        \multicolumn{7}{l}{\textbf{\textit{WavLM+ECAPA-TDNN (ours)}}}                                                                           \\
        +LDL+MVKL   & 4.48                         & 0.85                       & 1.68          & 6.17          & 0.84          & 1.70          \\
        +LDL+CVKL   & 4.19                         & 0.90                       & 1.59          & 5.95          & 0.85          & 1.67          \\
        +SVLDL+MVKL & 4.50                         & 0.84                       & 1.32          & 6.15          & 0.83          & 1.41          \\
        +SVLDL+CVKL & 4.16                         & \textbf{0.93}              & 1.34          & 5.91          & 0.86          & 1.43          \\
        ++Diff      & 4.19                         & 0.91                       & 1.08          & 5.93          & 0.85          & 1.07          \\
        +++Gender   & \textbf{4.14}                & 0.92                       & \textbf{1.03} & \textbf{5.82} & \textbf{0.87} & \textbf{1.05} \\ \bottomrule
    \end{tabular}
    \label{tab:result}
\end{table}

Ablation study results on the two datasets are reported in Table \ref{tab:result}.
When the backbone model is replaced by WavLM+ECAPA-TDNN, the backend model is still LDL+MVKL, and the performance of model regression is slightly improved.
If the backend model is replaced by CVKL, the performance of regression will be greatly improved.
If SVLDL is replaced, the multimodal problem can be solved to a certain extent.
This shows that under SVLDL, different variances are adaptively assigned to each utterance.
This results in the models not forcing them to optimize with the same variance, which also reduces multimodal problems.
Meanwhile, under CVKL, the model uses the CCC loss function, which can be directly optimized for the PCC (because the CCC is the unbiased PCC \cite{pandit2019many}), and the MAE is further reduced.
When the model simultaneously optimizes the L2 distance of the first-order difference of the age distribution, although the regression performance drops slightly, the multimodality problem is greatly solved.
When the auxiliary task of gender recognition is added, the model can distinguish gender information, thereby improving the age estimation performance.
Meanwhile, the multimodal problem is further solved.
Compared to the baseline model, the model achieves an MAE reduction of 10.39\% on NIST SRE08-10 and 6.58\% on the PA-Age dataset, outperforming the original model and close to the state-of-the-art model results.

\begin{figure}[ht]
    {\includegraphics[width=0.46\textwidth]{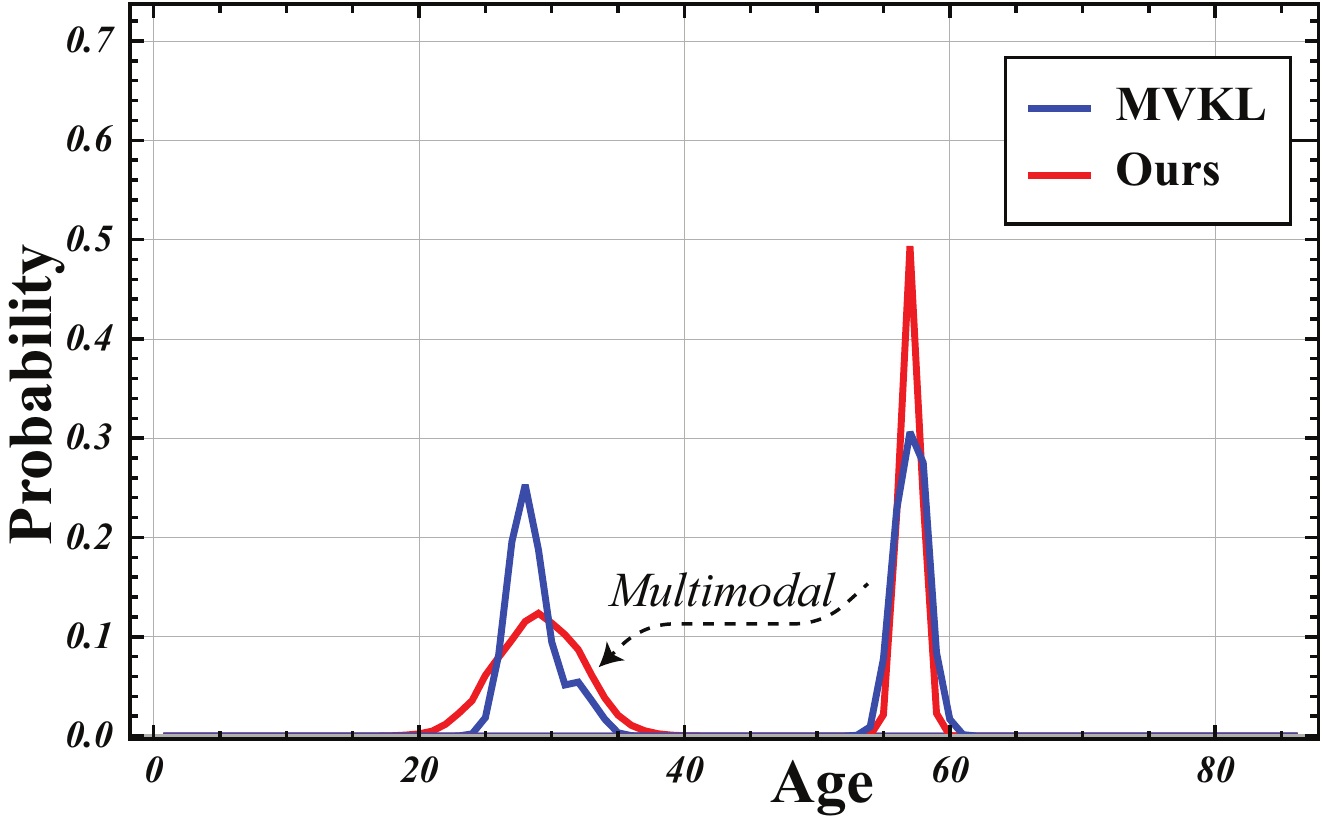}}
    \centering
    \caption{Distributions predicted by ``LDL + MVKL'' and ``SVLDL + CVKL + Diff + Gender'' (ours).
        The backbone model uses ``WavLM + ECAPA-TDNN''.
        Example from PA-Age dataset.}
    \label{fig:multimodal}
\end{figure}

As seen from Figure \ref{fig:multimodal}, our prediction is optimized to be unimodal and adaptively learned based on a specific instance, and has higher prediction performance.
That is, SVLDL adaptively selects variance for each instance to predict, Diff ensures unimodal distribution, and gender recognition improves prediction performance.
In contrast, MVKL is optimized for all instances with the same variance and cannot guarantee a unimodal distribution.

\begin{table}[ht]
    \centering
    \scriptsize
    \setlength{\tabcolsep}{4pt}
    \renewcommand{\arraystretch}{1.0}
    \caption{Ablation study of hyper-parameters for $ \lambda _{(2-5)} $ on PA-Age.
        The first row is when $ \lambda _4=\lambda _5=0 $, and the second row is when $ \lambda _2=1 $ and $ \lambda _3=0.1 $.}
    \begin{tabular}{@{}lcccccccccc@{}}
        \toprule
                             & MAE                             & $ \eta _2 $                           & MAE                                   & $ \eta _2 $                              & MAE                            & $ \eta _2 $   & MAE           & $ \eta _2 $   & MAE  & $ \eta _2 $ \\ \midrule
        $ \lambda _{(2,3)} $ & \multicolumn{2}{c}{(0.1, 0.01)} & \multicolumn{2}{c}{(0.1, 0.1)}        & \multicolumn{2}{c}{\textbf{(1, 0.1)}} & \multicolumn{2}{c}{(1, 1)}               & \multicolumn{2}{c}{(10, 1)}                                                                         \\
                             & 6.05                            & 1.63                                  & 5.97                                  & 1.49                                     & \textbf{5.91}                  & \textbf{1.43} & 6.12          & 1.54          & 6.47 & 1.50        \\ \midrule
        $ \lambda _{(4,5)} $ & \multicolumn{2}{c}{(0.01, 0)}   & \multicolumn{2}{c}{\textbf{(0.1, 0)}} & \multicolumn{2}{c}{(1, 0)}            & \multicolumn{2}{c}{\textbf{(0.1, 0.01)}} & \multicolumn{2}{c}{(0.1, 0.1)}                                                                      \\
                             & 5.98                            & 1.16                                  & \textbf{5.93}                         & \textbf{1.07}                            & 6.02                           & 1.07          & \textbf{5.82} & \textbf{1.05} & 5.88 & 1.06        \\ \bottomrule
    \end{tabular}
    \label{tab:hyperparam}
\end{table}

Table \ref{tab:hyperparam} shows the full path in the search for hyperparameter values of $ \lambda $(s) for the loss function, in Equation \ref{eq:loss}.
The first row shows the effect of adjusting $ \lambda _2 $ and $ \lambda _3 $ on predicted age (in MAE) and the multimodal parameter $ \eta _2 $ under $ \lambda _4=\lambda _5=0 $.
When $ \lambda _2=0.1 $, $ \lambda _3=0.01 $, the variance of the age distribution maybe too large, resulting in more multimodal problems.
Meanwhile, the prediction of age is not very accurate.
When $ \lambda_3 $ increases to 0.1, the variance becomes smaller, the age prediction becomes more accurate, and the multimodal problem is slightly solved.
The cases of $ \lambda_2=1 $ and $ \lambda_2=0.1 $ are the best-performing combination of $ \lambda_2 $ and $ \lambda_3 $.
But if these two hyperparameters are too large, it will affect the performance of age estimation, and the multimodal problem will appear again.

The second row in Table \ref{tab:hyperparam} is an experiment on $ \lambda _4 $ and $ \lambda _5 $ based on the best results from the first row (when $ \lambda_2=1 $ and $ \lambda_2=0.1 $).
When $ \lambda _5=0 $, the larger $ \lambda _4 $ is, the better the multimodal problem can be solved.
However, this case slightly degrades the age estimation performance.
Thus, we choose the best-performing case of $ \lambda _4=0.1 $, and adjust the value of $ \lambda _5 $.
After adding the auxiliary task of gender recognition, the performance of age estimation is further improved and achieves state-of-the-art results.
At the same time, the multimodal problem is further solved.
Therefore, the optimal hyperparameter combination is $ \lambda _1=10$, $\lambda _2=1 $, $ \lambda _3=\lambda _4=0.1 $ and $ \lambda _5=0.01 $.

\begin{table}[ht]
    \centering
    \small
    \setlength{\tabcolsep}{8pt}
    \renewcommand{\arraystretch}{1.0}
    \caption{Effect of duration of test utterance on PA-age estimated by our method compared to the best baseline results.}
    \begin{tabular}{@{}lllll@{}}
        \toprule
        MAE                                     & \multicolumn{4}{l}{Test segment lengths(s)}                                                 \\
        Model                                   & 10                                          & 15            & 20            & full          \\ \midrule
        ResNet-18+LDL+MVKL \cite{si2022towards} & 13.16                                       & 11.04         & 6.14          & 6.10          \\
        \textbf{Proposed method (ours)}         & \textbf{8.35}                               & \textbf{6.42} & \textbf{5.86} & \textbf{5.82} \\ \bottomrule
    \end{tabular}
    \label{tab:length}
\end{table}

Table \ref{tab:length} compares the effects of different test utterance durations on age estimation between the proposed method ``WavLM + ECPAP-TDNN + SVLDL + CVKL + Diff + Gender'' and the baseline model.
Here, only the utterances longer than 10 seconds are cut and selected as test data.
Compared to the baseline model, the proposed method achieves great improvement on short test utterances.
It may be because the WavLM speech features are trained based on big data, and the data augmentation method is used during training, which improves the robustness of the model.
At the same time, the pooling layer has an attention mechanism, which will make the model pay more attention to the information of speech and reduce over-fitting.

\begin{figure}[ht]
    \centering
    {\includegraphics[width=0.48\textwidth]{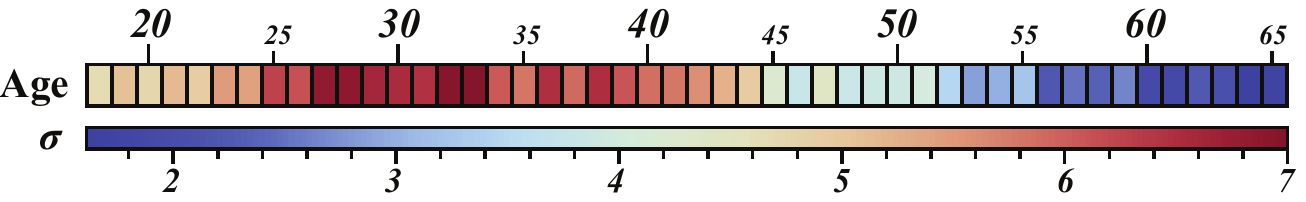}}
    \caption{The heat map to visualize the adaptively learned variances $ \sigma $ corresponding to different ages.}
    \label{fig:variance}
\end{figure}

Figure \ref{fig:variance} shows the median values of variance for adaptation at different ages.
It can be seen that before the age of 27, the voice is in a young and stable state.
From the age of 30 to 45, people's vocal cords gradually age, and the degree of aging varies from person to person, so the uncertainty is large.
After the age of 50, almost everyone's voice becomes older, the variance becomes smaller, and the vocal characteristics become more recognizable.

\section{Conclusions}
In this paper, a selective variance labeled distribution learning (SVLDL) method is proposed to accommodate variances of different age distributions.
The robustness of the age regression is enhanced by using Lin's consistent correlation coefficient (CCC) loss compared to the mean-variance-based loss.
Since existing methods suffer from multimodality in the age distribution, the quality of the fitted distribution is improved here by optimizing the L2 distance of the first-order difference of the distribution, and a reasonable proof is given.
The performance is further improved by using the speech features of WavLM and adding the auxiliary task of gender recognition.
Experiments show that the model achieves MAE reduction and multimodal problem-solving on both the NIST SRE08-10 and real-world PA-Age datasets, outperforming the original model to achieve state-of-the-art results.

\section{Acknowledgement}

This paper is supported by the Key Research and Development Program of Guangdong Province under grant No.2021B0101400003.
Corresponding author is Jianzong Wang from Ping An Technology (Shenzhen) Co., Ltd (jzwang@188.com).

\bibliographystyle{IEEEbib}

\bibliography{ref}

\end{document}